\documentclass[prd,superscriptaddress]{revtex4}

\usepackage{amsmath,amssymb}
\usepackage{verbatim}
\usepackage{graphicx}
\usepackage{hyperref}
\usepackage{color} 

\DeclareFontFamily{OT1}{rsfs}{}
\DeclareFontShape{OT1}{rsfs}{m}{n}{ <-7> rsfs5 <7-10> rsfs7 <10->rsfs10}{} 
\DeclareMathAlphabet{\mycal}{OT1}{rsfs}{m}{n} 

\begin{document}

\title{On two-vierbein gravity  action from gauge theory of conformal group}

\author{Iva Lovrekovic}
\email{ilovreko@.ic.ac.uk}
\affiliation{The Blackett Laboratory, Imperial College London,
Prince Consort Road, London SW7 2AZ, United Kingdom}

\date{\today}


\begin{abstract} 
We study the gravity action built from two gauge fields corresponding to the generators of the conformal group.  Starting with the action from which one can obtain Einstein gravity and conformal gravity upon imposing suitable constraints, we keep two independent gauge fields and integrate out the field corresponding to the generator of Lorentz transformations.
We identify the two gauge fields with two vierbeins and perturb them around an Anti--de Sitter space. This gives the linearized equations that differ from both, Einstein gravity and conformal gravity linearized equations. We also study the linearized equations  for one gauge field perturbed around the flat space and one around zero, and the case when the gauge fields are proportional to each other.
\end{abstract}


\maketitle


\section{Introduction}

Conformal gravity was interpreted as a gauge theory of conformal group O(4,2) by Kaku et. al \cite{Kaku:1977pa} in 1977. 
The motivation to study it was a fact that Einstein gravity has been viewed as a gauge theory of the de--Sitter group O(3,2) \cite{MacDowell:1977jt}, which upon contraction reduces to the Poincare group. Squaring the curvatures of de--Sitter group one obtains Einstein gravity \cite{MacDowell:1977jt}, while Poincare group and De--Sitter group are subgroups of the conformal group O(4,2).
It was natural to look at the square of the curvature of O(4,2). To achieve the invariance of constructed action under proper conformal gauge transformations the authors had to require that the gauge generator of translations vanishes.   Resulting action is invariant under conformal transformations and it is a gauge theory of the conformal group. It is built out of three independent gauge fields. Upon integrating out the gauge fields, we are left with the remaining two. This situation where one encounters two different fields appears in bimetric gravity models, which contain two dynamical metrics. 
These models  \cite{Hassan:2011zd, Hassan:2011ea, Hassan:2011hr} orginated from the dRGT massive gravity model \cite{deRham:2010ik,deRham:2010kj,deRham:2011rn,Paulos:2012xe}. It has been shown that other higher derivative theories, one of them being conformal gravity, can be rewritten and obtained from bimetric and partially massless bimetric theory \cite{Hassan:2013pca}. This has further motivated a study of bimetric gravity \cite{Hassan:2011zd}, whose  action  takes a form \cite{Hassan:2011zd}
\begin{align}
S=M_g^2\int d^4x \sqrt{-\text{det} g}R^{(g)}+M_f^2\int d^4x\sqrt{-\text{det} f} R^{(f)} +2m^2M_{\text{eff}}^2\int d^4x\sqrt{-\text{det} g}\sum_{n=0}^4\beta_ne_n(\sqrt{g^{-1}f}).
\end{align}
$R^{(g)}$ and $R^{(f)}$ are Ricci scalars with respect to metrics $g_{\mu\nu}$ and $f_{\mu\nu}$, $M_g$ and $M_f$ are two different Planck masses and $M_{\text{eff}}$ is an effective Planck mass. The $e_n$ are elementary symmetric polynomials in eigenvalues of $\sqrt{g^{-1}f}$, and $\beta_n$ are four combinations of the mass of the graviton, the cosmological constant and free parameters. The graviton mass and cosmological constants for $g_{\mu\nu}$ and $f_{\mu\nu}$ are among five free parameters of the theory.
  Four dimensional spin-2 theories have recently been studied within the different dimensional reduction schemes coming from 5-dimensional Chern-Simons gauge theories. The resulting actions were four dimensional generalizations of Einstein-Cartan theory, conformal gravity and bimetric gravity \cite{Albornoz:2018uin}.

Here, we study the linearized gravity, perturbed around maximally symmetric space, as a gauge theory of conformal group, while keeping two dynamical gauge fields. We find that perturbing the equations around AdS space gives degeneracy in the fields. The reason for this comes from the symmetric appearance of the gauge fields in the initial action and perturbation around maximally symmetric space. The linearized theory is different from the sum of linearized Einstein gravities for two metrics since the equations of motions do not come from corresponding Einstein actions, where linearized MacDowell-Mansouri action has been studied in \cite{Basile:2015jjd}. It also differes from the linearized conformal gravity since we do not require invariance under the proper conformal gauge transformations, and vanishing of the generator of translations which has in \cite{Kaku:1977pa} been imposed "by hand".

Comparison with the linearized EG and CG further shows that the original action should consist out of the two Ricci scalars, one for each metric, and an additional potential. Just like CG, action has one dimensionless parameter $\alpha$, but two dynamical gauge fields as one would expect from gauge theory for bimetric gravity. We also compare the linearized equations to linearized equations of bimetric gravity. 
One could remove the degeneracy between the fields by introducing a parameter multiplying one of the gauge fields, however the fields would still be linearly dependent. In order for them not to be linearly dependent one would need to have kinetic part modified.  Another possibility for removing the degeneracy would be to perturb the fields around different backgrounds, for example, one of the fields could be perturbed around AdS background and another around a black hole.
For now, we focus on the perturbations of both of the fields around AdS space, perturbation of one field around AdS space and the other around flat space, and on non-perturbative case where gauge fields depend linearly on each other. The content of the article is as follows.
The second section describes the action and corresponding equations of motion, while the third section analyzes them as a perturbation around the maximally symmetric spaces. In the fourth section we obtain the linearized equations of motion for the two gauge field fluctuations, perturbed around the AdS space. 
In section five we show example of the linearization around Minkowski space, while in the section six we consider the case when the gauge fields are proportional to each other. In section seven we discuss the results and possible future prospects.

\section{Action}
The most general parity conserving quadratic action that can be constructed using the  curvatures of conformal group, with no dimensional constants is \cite{Kaku:1977pa}
\begin{align}
I=\frac{\alpha}{8}\int d^4x \epsilon^{\mu\nu\rho\sigma}\epsilon^{abcd}R_{\mu\nu ab}(J)R_{\rho\sigma c d}(J)\label{actionoriginal}
\end{align}
for $\alpha$ dimensionless constant,
\begin{align}R_{\mu\nu ab}(J)&=\mathcal{R}_{\mu\nu ab}-2(e_{a\mu} f_{b\nu}-e_{b\mu}f_{a\nu})+2(e_{a\nu}f_{b\mu}-e_{b\nu} f_{a\mu}) \label{curvs1},
\end{align}
 and 
\begin{align} \mathcal{R}_{\mu\nu ab}=-\partial_{\mu}\omega_{\nu ab}+\partial_{\nu}\omega_{\mu ab}+\omega_{\mu a}^c\omega_{\nu cb}-\omega_{\nu a}^c\omega_{\mu c b}.\label{riccitensor}
\end{align}
It consists of the gauge fields $e_{a\mu}$ and $f_{a\mu}$ which appear symmetrically in action, and spin-connection $\omega_{\mu ab}$. 
If we rewrite the action using (\ref{curvs1}) and omit the topological invariant, Gauss-Bonnet term $(\mathcal{R}_{\mu\nu ab}(\omega))^2$ the action becomes
\begin{equation}
I=\frac{\alpha}{8}\int d^4 x \epsilon^{\mu\nu\rho\sigma}\epsilon^{abcd}(-16\mathcal{R}_{\mu\nu ab}e_{c\rho} f_{d\sigma}+64 e_{a\mu}f_{b\nu}e_{c\rho}f_{d\sigma})=\frac{\alpha}{8}\int d^4 xL \label{gbc},
\end{equation}
which contains three independent fields $\omega_{\mu ab}, e_{a\mu},$ and $f_{a\mu}$. The fields $e_{a\mu},$ and $f_{a\mu}$ appear symmetrically in action, so we treat them on equal footing.
If one imposes the requrement that the action is invariant under proper conformal gauge transformations, one needs to require that the gauge generator of translations 
\begin{align}
R_{\mu\nu a}(P)=-(\partial_{\mu}e_{a\nu}-\omega_{\mu}{}^b{}_ae_{b\nu})+(\partial_{\nu}e_{a\mu}-\omega_{\nu}{}^b{}_{a}e_{b\mu})+(e_{a\mu}b_{\nu}-e_{a\nu}b_{\mu})
\end{align} vanishes.
This constraint on the generator determines the gauge field $\omega_{\mu ab}$ identified with spin--connection. 
The gauge field $b_{\nu}$ is a generator of dilatations and it does not appear in the action. 
The action (\ref{gbc}) is scale and proper conformal invariant for $\omega=\omega(e)$. Keeping this spin connection, one can also integrate out the non-propagating field $f_{a\mu}$  to obtain the 
\begin{align}
I=\frac{\alpha}{8}\int d^4xC_{\mu\nu ab}C_{\rho\sigma cd}\epsilon^{\mu\nu\rho\sigma}\epsilon^{abcd},
\end{align}
conformal gravity action, here $C_{\mu\nu ab}$ is Weyl tensor.\\
One more approach to consider action is without background expectation value for the field $f_{a\mu}$. One can integrate out $f_{a\mu}$ to obtain action that depends on $\omega_{\mu ab}$ and $e_{a\mu}$. The action would be non-unitary and similar to Weyl squared action but different from it since $\omega_{\mu ab}$ would be an independent field and not a function of $e_{a\mu}$. 


\subsection{Equations of motion}

Varying the Lagrangean under the action (\ref{gbc}) with respect to $\omega_{\mu ab}$, one obtains its equation of motion 
\begin{align}
\delta_{\omega} L&
= \left( -2e_{c\nu}\partial_{\rho}f_{d\sigma}+2e_{c\nu}\omega_{\rho}^k{}_{d}f_{k\sigma}-2f_{c\nu}\partial_{\rho}e_{d\sigma}+2f_{c\nu}\omega_{\rho}^k{}_{d}e_{k\sigma} \right)\epsilon^{\mu\nu\rho\sigma}\epsilon^{abcd}=0 \label{omeom}
\end{align}
in terms of the $e_{a\mu}$ and $f_{a\mu}$ gauge fields.
Since the fields $e_{a\mu}$ and $f_{a\mu}$ appear symmetrically, we can compute the equation of motion for one gauge field and know it for the other gauge field as well. 
If we assume that $e_{a\mu}$ is invertible and has non-zero determinant we can determine its equation of motion from variation with respect to $e^{i\kappa}$
\begin{equation}
\delta_e L=\epsilon^{\mu\nu\kappa\sigma}\epsilon^{abid}\left[ -\mathcal{R}_{\mu\nu ab}f_{d \sigma}+8f_{b\nu}f_{d\sigma}e_{a\mu} \right]=0
\label{feom1},
\end{equation}
while for the analogous equation for $f^{i\kappa}$ we have to take analogous assumptions for $f_{a\mu}$
\begin{equation}
\delta_f L=\epsilon^{\mu\nu\kappa\sigma}\epsilon^{abid}\left[ -\mathcal{R}_{\mu\nu ab}e_{d \sigma}+8e_{b\nu}e_{d\sigma}f_{a\mu} \right]=0\label{feom2}
\end{equation}
which corresponds to \cite{Kaku:1977pa}
\begin{equation}
f_{ a\mu}=-\frac{1}{4}(R_{a \mu  }-\frac{1}{6}R e_{a\mu}).\label{feom}
\end{equation}
Here, we have used the contractions \begin{align}R_{b\mu }=\mathcal{R}_{\mu\nu ab}e^{a\nu}, && R=R_{a\mu }e^{a\mu}\label{eq28o}\end{align} and $f_{\mu\nu}=e^{a}{}_{\mu}f_{a\nu}$. Equation (\ref{feom2}) inserted back is known to give conformal gravity action for vanishing of translation generator \cite{Kaku:1977pa,Grumiller:2013mxa}\footnote{One can draw the analogous conclusion for the equation (\ref{feom1}).}. 
However, we keep both of the gauge fields dynamical and perturbatively solve equation (\ref{omeom}) for $\omega_{\mu ab}$.

We introduce pertrubations of the gauge fields 
\begin{align}
e_{\mu}^a&=v^{a}_{\mu}+\eta\chi^{a}_{\mu}+\eta^2\zeta^{a}_{\mu}+... \label{expe} \\
f_{\mu}^a&=f^{(0)a}_{\mu}+\eta\theta^{a}_{\mu}+\eta^2\psi^{a}_{\mu}+... \label{expf}
\end{align}
 and the perturbation of the spin-connection $\omega_{\mu ab}$ 
\begin{align}
\omega_{\mu ab}=\omega^{(0)}_{\mu ab}+\eta \omega^{(1)}_{\mu ab}+\eta^{2}\omega^{(2)}_{\mu ab}+...
\end{align}with $\eta$ small perturbation parameter.
In the expansion of curvatures in (\ref{feom2}) 
\begin{align}
R_{b\mu}=R^{(0))}_{b\mu}+\eta R^{(1)}_{b\mu}+...
\end{align}
for $R^{(0)}_{b\mu}=\mathcal{R}^{(0)}_{\mu\nu ab}v^{a\nu}$, one needs to take into account the contractions $R^{(1)}_{b\mu}=\mathcal{R}^{(1)}_{\mu\nu ab}v^{a\nu}+\mathcal{R}^{(0)}_{\mu\nu ab}\tilde{\chi}^{a\nu}$ from (\ref{eq28o}).
Analogously, the expansion of the Ricci scalar is
\begin{align}
R=R^{(0)}_{b\mu}v^{b\mu}+\eta (R^{(1)}_{b\mu}v^{b\mu}+R^{(0)}_{b\mu}\tilde{\chi}^{b\mu})+...
\end{align}

The allowed vacuum points around which we can perturb the action and equations of motion, need to be backgrounds with curvature. One would naively
 perturb the fields around the flat background, however the choice of $e_{a\mu}=f_{a\mu}$ would not satisfy equation of motion for $f_{a\mu}$ or $e_{a\mu}$ if both of them are flat. If one of them was flat, the other one would have to be zero. One could further analyze around which backgrounds is it allowed to perturb the solution by studying the allowed solutions, as it was done for the Einstein theory  in \cite{Tseytlin:1981ks}.

\section{Perturbation around $v^{}_{a\mu}=f^{(0)}_{a\mu}$}

We choose the background with $v^{a}_{\mu}=f^{(0)a}_{\mu}$.
In the leading order the solution for the equation of motion (\ref{omeom}) is 
\begin{equation}
\omega^{(0)}_{\nu ab}=-\tfrac{1}{2}  (v^{}_{b}{}^{\beta} \partial_{\beta}v_{a\nu} + v^{}_a{}^{\alpha}v_b^{\beta}v_{\nu}^{c}(-\partial_{\alpha}v^{}_{c\beta}+\partial_{\beta}v^{}_{c \alpha}) -  v^{}_{a}{}^{\beta} \partial_{\beta}v^{}_{b\nu} -  v^{}_{b}{}^{\beta} \partial_{\nu}v^{}_{a\beta} + v^{}_{a}{}^{\beta} \partial_{\nu}v^{}_{b\beta})\label{specom}
\end{equation}
which agrees with the well known spin-connection for Einstein gravity. The leading order equations (\ref{feom1}) and (\ref{feom2}) will expectedly give equal solution, which is Einsten action with the cosmological constant 
\begin{align}
R^{(0)}_{\mu\nu}-4v^{}_{\mu\nu}=0. \label{lo}
\end{align}
Here we have defined $v^{}_{\mu\nu}=v^{}_{b\mu}v^{b}{}_{\nu}$.
 For the analysis of the linear order it is convenient to introduce the tensor 
\begin{align}
e_{a\mu}f_{b\nu}=Q_{ab\mu\nu},
\end{align}whose subleading order reads \begin{align}
Q^{(1)}_{ab\mu\nu}=v_{b\nu}\chi_{a\mu}+v_{a\mu}\theta_{b\nu},
\end{align}
and rewrite the subleading order of the equation (\ref{omeom}) in terms of it
\begin{align}
v^{d}{}_{[\nu}v^{k}{}_{\sigma}\omega^{(1)}_{\rho]}{}^{c}{}_{k}-v^c{}_{[\nu}v^{k}{}_{\sigma}\omega^{(1)}_{\rho]}{}^{d}{}_k-\partial_{[\rho}Q^{(1)[cd]}{}_{\nu\sigma]}=0.\label{qeom}
\end{align}
The combinations of the $Q_{ab\mu\nu}$ tensor which appear in the equation (\ref{qeom}) allow rewriting the partial derivatives in terms of the general covariant derivative defined on the background space, because the Christoffels and spin-connections added and subtracted to form the covariant derivative, exactly cancel. One obtains\begin{align}
\omega^{(1)}_{\kappa ab}=\frac{1}{2}v_c{}^{\alpha}v_{d}{}^{\beta}(v_{b\kappa}v_{a}{}^{\gamma}-v_b{}^{\gamma}v_{a\kappa})\mathcal{\nabla}_{[\alpha}Q^{(1)[cd]}{}_{\beta\gamma]}+v_d{}^{\beta}(v_a{}^{\alpha}\eta_{bc}-v_{b}{}^{\alpha}\eta_{ac})\mathcal{\nabla}_{[\alpha}Q^{(1)[cd]}{}_{\delta\kappa]}.\label{qomega}
\end{align}
The subleading order of the spin connection consists of the background vielbeins  which are solution of (\ref{lo}), Einstein spaces, and fluctuations $\chi_{a\mu}$, $\theta_{a\mu}$, which will be defined through (\ref{feom1}) and (\ref{feom2}). 
The subleading order of (\ref{feom1})
\begin{align}\
\theta^{}_{b\mu}=-\frac{1}{4}(R^{(1)}_{b\mu}-\frac{1}{6}R^{(0)}\chi_{b\mu}-\frac{1}{6}R^{(1)}v^{}_{b\mu})\label{eq53},
\end{align}
consists of
\begin{align}
R^{(1)}_{b\mu}&=\left(-\partial_{\mu}\omega_{\nu ab}^{(1)}+\partial_{\nu}\omega_{\mu ab}^{(1)}+\omega^{c(0)}_{\mu a}\omega^{(1)}_{\nu c b}-\omega^{c(0)}_{\nu a}\omega_{\mu cb}^{(1)}+\omega^{c(1)}_{\mu a}\omega^{(0)}_{\nu c b}-\omega^{c(1)}_{\nu a}\omega_{\mu cb}^{(0)}\right)v^{a\nu}\nonumber\\&+(-\partial_{\mu}\omega^{(0)}_{\nu ab}+\partial_{\nu}\omega_{\mu ab}^{(0)}+\omega^{c(0)}_{\mu a}\omega^{(0)}_{\nu c b}-\omega^{c(0)}_{\nu a}\omega_{\mu cb}^{(0)})\tilde{\chi}^{a\nu}\label{eq54},
\end{align} for
\begin{align}R^{(1)}=R^{(1)}_{b\mu}v^{b\mu}+R^{(0)}_{b\mu}\tilde{\chi}^{b\mu}\label{eq55}\end{align}
and $R^{(0)}=R^{(0)}_{b\mu}e^{b\mu},$  and gives the dependence of $\chi_{a\mu}$ and $\theta_{a\mu}$.

\section{AdS Background}

We set the background perturbation to the AdS metric, which is Weyl flat allowing us to write
\begin{align}
v^{}_{a\mu}=\rho(x)\delta_{a\mu}
\end{align}
and the leading order spin-connection
\begin{align} 
\omega^{(0)}_{\nu ab}=-\delta_{[a\nu}\partial_{b]}\rho(x),
\end{align}
here we denote, $\partial_b=\delta_b^{\mu}\partial_\mu$.   The equations (\ref{feom1}) and (\ref{feom2})  reduce to $R^{(1)}_{a\mu}=4\delta_{\mu\nu}$.
The subleading order of the equation (\ref{omeom}), just as (\ref{qomega}) after few technical manipulations, shows that linear term in the $\omega_{\mu ab}$  perturbation can be rewritten in terms of the sum of two linear terms of Einstein spin connections
\begin{align}
\omega^{(1)}_{\kappa ak}=\omega^{(1)}_{\kappa ak}(\chi)+\omega^{(1)}_{\kappa ak}(\theta)\label{omegap1}
\end{align}
Here, \begin{align}
\omega^{(1)}_{\kappa ak}(\chi)=-\frac{1}{4\rho}\left( \delta^{\alpha}_a\nabla_{\alpha}\chi_{k\kappa}+\delta^{\alpha}_k\nabla_{\kappa}(\chi^{}_{a\alpha})+\delta^{\alpha}_k\delta^{b}_{\kappa}\delta^{\beta}_a\nabla_{\beta}\chi^{}_{b\alpha} \right)-a \leftrightarrow k
\end{align}
is linearized spin connection for Einstein gravity, and $\nabla$ denotes Lorentz covariant derivative. For transparency, we keep the Lorentz covariant derivative, and do not evaluate it for background AdS. The expression for the linearized spin connection evaluated on AdS is given in the appendix.
This form of $\omega^{(1)}_{\mu ab}$ allows to split the curvatures in parts depending only on $\chi_{a\mu}$ or $\theta_{a\mu}$ fluctuation. 
Therefore, we can write the subleading order of the Riemann tensor as sum of linearized Riemann tensors for Einstein gravity. Subleading order of Ricci tensor however will not be possible to write in the form of the two linearized Ricci tensors for Einstein gravity, because of the term $R^{(0)}_{\mu\nu ab}\tilde{\chi}^{a\nu}$ ($R^{(0)}_{\mu\nu ab}\tilde{\theta}^{a\nu}$) visible from (\ref{eq54})
\begin{align}
R_{b\mu}^{(1)}=(R^{(1)}_{\mu\nu ab}(\chi)+R^{(1)}_{\mu\nu ab}(\theta))v^{a\nu}+R^{(0)}_{\mu\nu ab}\tilde{\chi}^{a\nu}\label{eq67nn}
\end{align}
Here 
\begin{align}
R^{(1)}_{\mu\nu ab}(\chi)=-\partial_{\mu}\omega_{\nu ab}^{(1)}(\chi)+\partial_{\nu}\omega_{\mu ab}^{(1)}(\chi)+\omega^{c(0)}_{\mu a}\omega^{(1)}_{\nu c b}(\chi)-\omega^{c(0)}_{\nu a}\omega_{\mu cb}^{(1)}(\chi)+\omega^{c(1)}_{\mu a}(\chi)\omega^{(0)}_{\nu c b}-\omega^{c(1)}_{\nu a}(\chi)\omega_{\mu cb}^{(0)}\label{eq72n},
\end{align}
is linearized Riemann tensor for Einstein gravity. 
We contract the equation (\ref{eq53}) with $v^{b}{}_{\sigma}$ and write 
\begin{align}\
\theta^{}_{b\mu}v^{b}{}_{\sigma}=-\frac{1}{4}(R^{(1)}_{b\mu}v^{b}{}_{\sigma}-\frac{1}{6}R^{(0)}\chi_{b\mu}v^{b}{}_{\sigma}-\frac{1}{6}R^{(1)}v^{}{}_{b\mu}v^{b}{}_{\sigma})\label{eq119n}.
\end{align}
\noindent In terms of the Einstein gravity perturbations in the fields $\chi_{a\mu}$ and $\theta_{a\mu}$, using (\ref{eq67nn}) and (\ref{eq55}), this is 
\begin{align}
\theta^{}_{b\mu}v^{b}{}_{\sigma}&=-\frac{1}{4}((R^{(1)}_{\mu\nu ab}(\chi)+R^{(1)}{}_{\mu\nu ab}(\theta)v^{a\nu}+R^{(0)}{}_{\mu\nu ab}\tilde{\chi}^{a\nu}-\frac{1}{6}R^{(0)}\chi^{}_{b\mu})v^{b}{}_{\sigma} \label{thcon}
\\&-\frac{1}{6}((R^{(1)}_{\alpha\nu ac}(\chi)+R^{(1)}_{\alpha\nu ac}(\theta))v^{a\nu}+R^{(0)}_{\alpha\nu ac}\tilde{\chi}^{a\nu})v^{c\alpha}v^{}_{b\mu}v^{b}{}_{\sigma}-\frac{1}{6}R^{(0)}_{c\alpha}\tilde{\chi}^{c\alpha}v^{}_{b\mu}v^{b}{}_{\sigma})\nonumber.
\end{align}
This way one obtains the constraint on the $\chi_{\mu\nu}$ related to $\theta_{\mu\nu}$. 
Analogous appearance of both equations of motions for $f_{a\mu}$ and $e_{a\mu}$ gauge fields assuming them invertible implies that equation for $\chi_{b\mu}$ is
\begin{align}
\chi^{}_{b\mu}v^{b}{}_{\sigma}&=-\frac{1}{4}((R^{(1)}_{\mu\nu ab}(\theta)v^{a\nu}+R^{(1)}{}_{\mu\nu ab}(\chi)v^{a\nu}+R^{(0)}{}_{\mu\nu ab}\tilde{\theta}^{a\nu}-\frac{1}{6}R^{(0)}\theta^{}_{b\mu})v^{b}{}_{\sigma} \label{chiconx}
\\&-\frac{1}{6}(R^{(1)}_{\alpha\nu ac}(\theta)v^{a\nu}v^{c\alpha}+R^{(1)}_{\alpha\nu ac}(\chi)v^{a\nu}v^{c\alpha}+2R^{(0)}_{\alpha\nu ac}\tilde{\theta}^{a\nu}v^{c\alpha})v^{}_{b\mu}v^{b}{}_{\sigma})\nonumber. 
\end{align}

\noindent If we subtract the equations (\ref{chiconx}) and (\ref{thcon}) we obtain
\begin{align}
(\theta^{}_{b\mu}-\chi^{}_{b\mu})v^{b}{}_{\sigma}&=-\frac{1}{4}((R^{(0)}{}_{\mu\nu ab}(\tilde{\chi}^{a\nu}-\tilde{\theta}^{a\nu})-\frac{1}{6}R^{(0)}(\chi^{}_{b\mu}-\theta^{}_{b\mu}))v^{b}{}_{\sigma} 
\\&-\frac{1}{6}(2R^{(0)}_{\alpha\nu ac}v^{c\alpha})v^{}_{b\mu}v^{b}{}_{\sigma}(\tilde{\chi}^{a\nu}-\tilde{\theta}^{a\nu}))\nonumber. 
\end{align} The equation does not contain any lilnearized curvatures due to their cancellation. The reason for this is that the terms with the linearized Riemann tensor can be written as a sum of the linear Riemann tensor for Einstein gravity and contain both perturbations, $\chi_{a\mu}$ and $\theta_{a\mu}$ in each equation (34) and (35). Subtracting the equations will cancel these terms.
\noindent Using the conventions $R^{(0)}_{\mu\nu\alpha\beta}=-\tilde{\lambda}(-v^{}_{\mu\beta}v^{}_{\nu\alpha}+v^{}_{\mu\alpha}v^{}_{\nu\beta})$, $R_{\alpha\beta}=3\tilde{\lambda}v^{}_{\alpha\beta}$, and $\tilde{\chi}^{a\nu}=-\chi^{a\nu}$ and  $\tilde{\theta}^{a\nu}=-\theta^{a\nu}$ we evaluate (36) and get
\begin{align}
\tilde{\lambda}(\theta_{\mu\sigma}-\chi_{\mu\sigma})=2(2+\tilde{\lambda})(\theta_{\sigma\mu}-\chi_{\sigma\mu})\label{eq50n}
\end{align}
for $\tilde{\lambda}=-1$ this is
\begin{align}
\chi_{\mu\sigma}-\theta_{\mu\sigma}=2(\theta_{\sigma\mu}-\chi_{\sigma\mu})\label{eq51n}
\end{align}
or
\begin{align}
\theta_{\mu\sigma}+2\theta_{\sigma\mu}&=2\chi_{\sigma\mu}+\chi_{\mu\sigma}\label{constr1}.
\end{align}
Due to Lorentz invariance, we can impose a gauge that $\chi_{a\mu}$ is symmetric matrix, $\chi_{a\mu}=\chi_{\mu a}$. This would imply that $\chi_{a\mu}v^{a}_{\nu}=\chi_{\mu a}v^{a}_{\nu}\rightarrow\chi_{\mu\nu}=\chi_{\nu\mu}$. This condition requires 
\begin{align}
\theta_{\mu\sigma}+2\theta_{\sigma\mu}=3\chi_{\sigma\mu}.\label{constraint1}
\end{align}
\noindent Summing the equations (\ref{chiconx}) and (\ref{thcon}) and using the same notation gives
\begin{align}
(\theta^{}_{\sigma\mu}+\chi^{}_{\sigma\mu})=-\frac{1}{4}(k.t.+\tilde{\lambda}(\chi^{}_{\mu\sigma}+\theta^{}_{\mu\sigma})-2\tilde{\lambda}(\chi^{}_{\sigma\mu}+\theta^{}_{\sigma\mu}))\label{sumcon}
\end{align}
for  \textit{k.t.} kinetic term
\begin{align}
k.t.&=(2R^{(1)}_{\mu\nu ab}(\chi+\theta)v^{a\nu}v^{b}{}_{\sigma}-\frac{1}{3}R^{(1)}_{\alpha\nu ac}(\chi+\theta)v^{a\nu}v^{c\alpha}v^{}_{\mu\sigma}).\end{align}
To evaluate the linear term $R^{(1)}_{\mu\nu ab}=\delta R_{\mu\nu ab}$ we linearize the tensor in the metric formulation and use projection to the tetrad formulation
\begin{align}
\delta R^{}_{\mu\nu cd}(\chi)\equiv R^{(1)}_{\mu\nu cd}(\chi)= R^{(1)}{}_{\lambda\sigma \mu\nu}v_{c}^{\lambda}v_d^{\sigma}(\chi)-R^{(0)}_{\mu\nu ab}\delta^{a}_c\chi^{b}{}_{d}-R^{(0)}_{\mu\nu ab}\chi^{a}{}_c\delta^{b}_{d} \label{eq43}
\end{align}
we obtain
\begin{align} \label{eq44n}
k.t.&=6\tilde{\lambda}(h_{\mu\sigma}+q_{\mu\sigma})-\mathcal{D}_{\sigma}\mathcal{D}_{\mu}(h+q)-\mathcal{D}^2(h_{\mu\sigma}+q_{\mu\sigma})+2\mathcal{D}_{(\mu}\mathcal{D}_{\alpha}(h_{\sigma)}^{\alpha}+q_{\sigma)}^{\alpha})\\&-\frac{1}{3}(3\tilde{\lambda}(h+q)-\mathcal{D}^2(h+q)+\mathcal{D}_{\alpha}\mathcal{D}_{\beta}(h^{\alpha\beta}+q^{\alpha\beta}))v_{\mu\sigma}-2\tilde{\lambda}(\chi_{\mu\sigma}+\theta_{\mu\sigma}). \nonumber
\end{align}
Here, we have defined $h_{\mu\nu}=v_{a\mu}\chi^a{}_{\nu} +v_{a\nu}\chi^a{}_{\mu}$ and $q_{\mu\nu}=v_{a\mu}\theta^a{}_{\nu}+v_{a\mu}\theta^a{}_{\nu}$, their traces $h$ and $q$ respectively, and we 
 have not used any gauge conditions.  The last term in (\ref{eq44n}) comes from the two last terms in (\ref{eq43}).
 For the sum of the constraint equations on the linear term in perturbation of gauge field, from (\ref{sumcon}) we obtain 
\begin{align}
0&=-\mathcal{D}_{\sigma}\mathcal{D}_{\mu}(h+q)-\mathcal{D}^2(h_{\mu\sigma}+q_{\mu\sigma})+2\mathcal{D}_{(\mu}\mathcal{D}_{\alpha}(h_{\sigma)}^{\alpha}+q_{\sigma)}^{\alpha}) \nonumber \\&-\frac{1}{3}(-\mathcal{D}^2(h+q)+\mathcal{D}_{\alpha}\mathcal{D}_{\beta}(h^{\alpha\beta}+q^{\alpha\beta}))v_{\mu\sigma}+6\tilde{\lambda}(h_{\mu\sigma}+q_{\mu\sigma})-\tilde{\lambda}(h+q)v_{\mu\sigma}\nonumber\\&-\tilde{\lambda}(\chi_{\mu\sigma}+\theta_{\mu\sigma})-(2\tilde{\lambda}-4)(\chi_{\sigma\mu}+\theta_{\sigma\mu})
\end{align}
for $2\mathcal{D}_{(\mu}\mathcal{D}_{\alpha}h_{\sigma)}{}^{\alpha}= \mathcal{D}_{\mu}\mathcal{D}_{\alpha}h_{\sigma}{}^{\alpha}+\mathcal{D}_{\sigma}\mathcal{D}_{\alpha}h_{\mu}{}^{\alpha}$.
One can also choose the De Donder gauge $\mathcal{D}_{\alpha}(h^{\alpha}{}_{\beta}+q^{\alpha}{}_{\beta})=\frac{1}{2}\mathcal{D}_{\beta}(h+q)$ which keeps in the equation Laplace operators acting on the sum of the symmetrized linear terms in expansion of gauge field, their traces and mass terms
\begin{align}
0&=-\mathcal{D}^2(h_{\mu\sigma}+q_{\mu\sigma}) \nonumber +\frac{1}{6}\mathcal{D}^2(h+q)v_{\mu\sigma}+6\tilde{\lambda}(h_{\mu\sigma}+q_{\mu\sigma})-\tilde{\lambda}(h+q)v_{\mu\sigma}\nonumber\\&-\tilde{\lambda}(\chi_{\mu\sigma}+\theta_{\mu\sigma})-(2\tilde{\lambda}-4)(\chi_{\sigma\mu}+\theta_{\sigma\mu})
\label{eq56}
\end{align}
For $\tilde{\lambda}=-1$ (\ref{eq56}) becomes
\begin{align}
v^a{}_{\sigma}T^{(1)}{}_{a\mu}&\equiv-\mathcal{D}^2(h_{\mu\sigma}+q_{\mu\sigma}) \nonumber +\frac{1}{6}\mathcal{D}^2(h+q)v_{\mu\sigma}-6(h_{\mu\sigma}+q_{\mu\sigma})+(h+q)v_{\mu\sigma}\nonumber\\&+(\chi_{\mu\sigma}+\theta_{\mu\sigma})+6(\chi_{\sigma\mu}+\theta_{\sigma\mu})\nonumber\\&=-\mathcal{D}^2(h_{\mu\sigma}+q_{\mu\sigma})  +\frac{1}{6}\mathcal{D}^2(h+q)v_{\mu\sigma}-5(h_{\mu\sigma}+q_{\mu\sigma})+5(\chi_{\sigma\mu}+\theta_{\sigma\mu})+(h+q)v_{\mu\sigma}
\label{eqx}
\end{align}
 where we call this equation $v^a{}_{\sigma}T^{(1)}{}_{a\mu}$.
From the equations (\ref{eq51n}) and (\ref{eqx})  one can notice that fluctuations cannot be fixed independently, they appear as a sum, which implies that here is an extra symmetry.

Highly symmetric equations (\ref{eq51n}) and (\ref{eqx}) are pointing out the degeneracy of the perturbations around the maximally symmetric background. This becomes obvious when one tries to symmetrize equation (\ref{eq51n}). 
One obtains the equality $\chi_{\mu\sigma}+\chi_{\sigma\mu}=\theta_{\sigma\mu}+\theta_{\mu\sigma}$  which inserted into symmetrized equation (\ref{eqx}) 
leads to  two equal equations for $\chi_{\mu\sigma}+\chi_{\sigma\mu}$ and $\theta_{\mu\sigma}+\theta_{\sigma\mu}$.  One could further analyze symmetrized equation (\ref{eqx})
\begin{align}
0&=-2\mathcal{D}^2(h_{\mu\sigma}+q_{\mu\sigma})  +\frac{1}{3}\mathcal{D}^2(h+q)v_{\mu\sigma}-5(h_{\mu\sigma}+q_{\mu\sigma})+2(h+q)v_{\mu\sigma}
\label{eqxsym}
\end{align}
 rewritng the perturbations in the transverse traceless split, and consider its one loop partition function, however, one would have to keep in mind the implications of the equations (\ref{eq51n}).
\\
\indent Independently, one can antisymmetrize equation (\ref{eqx}) which will lead to cancellation of derivatives and $\chi_{\mu\sigma}-\chi_{\sigma\mu}=-\theta_{\mu\sigma}+\theta_{\sigma\mu}$. With Lorentz invariance requirement that $\chi_{\sigma\mu}$ is symmetric, antisymmetrizing (\ref{constraint1}) one obtains that $\theta_{\sigma\mu}$ is also symmetric. (\ref{constraint1}) will then lead to $\theta_{\mu\sigma}=\chi_{\mu\sigma}$.\\
The equation (\ref{eqxsym}) however can't be compared to the known linearized equations of EG or CG. As shown in the Appendix C on the example of Einstein gravity, projection of general perturbed tensor $T_{\mu\nu}=T^{(0)}_{\mu\nu}+\eta T^{(1)}_{\mu\nu}$ is $T^{(1)}_{\mu\nu}=v^a{}_{\mu}T^{(1)}{}_{a\nu}+\chi^a{}_{\nu}T^{(0)}_{ a\mu}$. We can recognize (\ref{eqx}) as $v^a{}_{\mu}T^{(1)}{}_{a\nu}$
part of the equation. To be able to compare the equation with linearized EG and CG from the literature we have to obtain $T^{(1)}_{\mu\nu}$, i.e. we have to add  $\chi^a{}_{\nu}T^{(0)}_{a\mu }$ to $v^a{}_{\mu}T^{(1)}{}_{a\nu}$ tensor. 
After that, equation (\ref{eqx}) becomes  
\begin{align}
T^{(1)}_{\sigma\mu}=-8(\chi_{\sigma\mu}+\theta_{\sigma\mu})-5(h_{\mu\sigma}+q_{\mu\sigma})-\mathcal{D}^2(h_{\mu\sigma}+q_{\mu\sigma})-v_{\mu\sigma}(-(h+q)-\frac{1}{6}\mathcal{D}^2(h+q))=0\label{eq49}
\end{align}
which can be symmetrized to give 
\begin{align}
-9(h_{\mu\sigma}+q_{\mu\sigma})-\mathcal{D}^2(h_{\mu\sigma}+q_{\mu\sigma})-v_{\mu\sigma}(-(h+q)-\frac{1}{6}\mathcal{D}^2(h+q))=0.\label{eq50}
\end{align}
One can compare this to the linearized minimal bimetric gravity model where for the massless spin-2 particle $h_{\mu\nu}$ and a massive spin-2 particle $u_{\mu\nu}$ of mass $m$, one has
\cite{Hassan:2011zd}
\begin{align}
S=\int d^4x(h_{\mu\nu}\hat{\mathcal{\epsilon}}^{\mu\nu\alpha\beta}h_{\alpha\beta}+u_{\mu\nu}\hat{\mathcal{\epsilon}}^{\mu\nu\alpha\beta}u_{\alpha\beta})-\frac{m^2}{4}\int d^4x(u^{\mu\nu}u_{\mu\nu}-u^{\mu}{}_{\mu}u^{\nu}{}_{\nu}). \label{eq51}
\end{align}
Here, $\hat{\epsilon}^{\mu\nu\alpha\beta}$ denotes the Einstein-Hilbert (EH) kinetic operator.
One can notice that (\ref{eq49}) as well as linear equations that would come from  (\ref{eq51}), have the form of two equal operators acting on two separate fields  and a mass term. In (\ref{eq49}) the kinetic operator is not EH. One could think of the equation as consisted from two EH operators and additional mass terms. 
 When equation (\ref{eq49}) is symmetrized  and one obtains equation (\ref{eq50}), there are two equal kinetic operators for two degenerate fields, which can be thought as two EH operators and mass terms. Upon lifting the degeneracy between the fields, one should be able to diagonalize the resulting equation such that there are two EH operators, one for each field, and remaining terms which belong only to one massive field as in (\ref{eq51}).

 Analysis of the spin two massive graviton has been done in tetrad formulation for dRGT model using similar methods \cite{Mazuet:2018ysa}.
Possibly
convenient way for further considerations might be in terms of the field $Q_{\mu\nu\alpha\beta}$.
If we express the subleading order equation (\ref{eq51n})  in terms of this tensor, it reads
\begin{align}
Q^{(1)}_{\beta\mu\nu\sigma}-Q^{(1)}_{\nu\sigma\beta\mu}=2(Q^{(1)}_{\sigma\nu\mu\beta}-Q^{(1)}_{\mu\beta\sigma\nu}),
\end{align}
while symmetrized equation (\ref{eqxsym}) is
\begin{align}
0=&-2\mathcal{D}^2
(Q^{(1)}_{\mu\beta\sigma\nu}+Q^{(1)}_{\beta\mu\nu\sigma}+Q^{(1)}_{\sigma\nu\mu\beta}+Q^{(1)}_{\nu\sigma\beta\mu})+\frac{1}{3}\mathcal{D}^2Q^{(1)}v_{\mu\sigma}v_{\beta\nu}\nonumber\\&+2Q^{(1)}v_{\mu\sigma}v_{\beta\nu}-5(Q^{(1)}_{\mu\beta\sigma\nu}+Q^{(1)}_{\beta\mu\nu\sigma}+Q^{(1)}_{\sigma\nu\mu\beta}+Q^{(1)}_{\nu\sigma\beta\mu})
\end{align}
It can be useful to notice the property
\begin{align}
Q^{(1)}_{\nu\mu\beta\sigma}+Q^{(1)}_{\beta\sigma\nu\mu}=Q^{(1)}_{\beta\mu\nu\sigma}+Q^{(1)}_{\nu\sigma\beta\mu}
\end{align}

\section{$e_{a\mu}$ perturbed around the flat background, $f_{a\mu}$ around zero}

The linearized equations of motion when $e_{a\mu}$ is perturbed around the flat background and $f_{a\mu}$ around zero in (\ref{expe}) and (\ref{expf}) imply $\delta^a_{\mu}$ and zero respectively for leading order terms, and the subleading terms remain to be determined. 
The equation of motion for $\omega_{\mu ab}$ in the leading order vanishes because it is multiplied with leading order term in expansion of $f_{a\mu}$. This naturally makes (\ref{feom1}) and (\ref{feom2}) to be identically zero.\\
The subleading order of $\omega_{\mu ab}$  \begin{equation}
\omega^{(1)}_{\mu ab}=\tfrac{1}{6} \bigl(\delta_{b}{}^{\rho} (- \partial_{\mu}\chi_{a\rho} + \partial_{\rho}\chi_{a\mu}) + \delta_{a}{}^{\rho} (\partial_{\mu}\chi_{b\rho} -  \partial_{\rho}\chi_{b\mu}) + \delta_{a}{}^{\rho} \delta_{b}{}^{\alpha} e^{(0)d}{}_{\mu} (\partial_{\alpha}\chi_{d\rho} -  \partial_{\rho}\chi_{d\alpha})\bigr)\end{equation} agrees with a subleading term of $\omega_{\mu ab}$ in Einstein gravity,
while the subleading order of (\ref{feom2}) is
\begin{align}
\theta^{(1)}_{b\mu}=-\frac{1}{4}(R_{b\mu}^{(1)}-\frac{1}{6}R^{(1)}\delta_{b\mu}). 
\end{align}
The curvature terms in expansion are $R^{(1)}_{b\mu}=(-\partial_{\mu} \omega^{(1)}_{\nu ab}+\partial_{\nu} \omega^{(1)}_{\mu ab})\delta^{a\nu}$ and $R^{(1)}=R^{(1)}_{b\mu}\delta^{b\mu}$.
Following the procedure of the previous chapter
\begin{align}
R^{(1)}_{b\beta}=\frac{1}{2}(\partial_{\alpha}\partial_{\gamma}h^{\gamma}_{\beta}-\partial_{\beta}\partial_{\alpha}h+\partial_{\beta}\partial_{\gamma}h_{\alpha}^{\gamma}-\partial_{\gamma}\partial^{\gamma}h_{\alpha\beta})\delta^{\alpha}_b
\end{align}
and
\begin{align}
R^{(1)}=\partial_{\beta}\partial_{\alpha}h^{\alpha\beta}-\partial_{\beta}\partial^{\beta}h
\end{align}
Using the De Donder gauge and writting the  derivatives with $\mathcal{D}$ 
\begin{align}
\theta_{\sigma\mu}=\frac{1}{8}(\mathcal{D}^{\alpha}\mathcal{D}_{\alpha} h_{\sigma\mu}-\frac{1}{3}\mathcal{D}_{\alpha}\mathcal{D}^{\alpha}h\delta_{\sigma\mu}).\label{eq53nn}
\end{align}
 We can notice that there is dependency only on $\chi$ on the right hand side of (\ref{eq53nn}), which is a result of the fact that in the $\omega^{(1)}_{\mu ab}$ we have only $\chi_{a\mu}$ appearing. The subleading order of $\omega^{(1)}_{\mu ab}$ does not depend on $\theta_{a\mu}$ because in equation of motion that determines $\omega^{(1)}_{\mu ab}$, fields $f_{a\mu}$ appear in pairs, which will make such terms vanish in subleading order for the $f^{(0)}_{a\mu}$ vanishing. 

The leading order of the second equation (\ref{feom1}), will vanish because the perturbation of $f_{a\mu}$ field is expanded around zero. The subleading order will also vanish because the first term of (\ref{feom1}) is given by $R^{(0)}_{\mu\nu ab}\theta_{a\mu}+R^{(1)}_{\mu\nu ab}f^{(0)}_{a\mu}$ both of which vanish. The second term in (\ref{feom1}) will have multiplication with vanishing background $f^{(0)}_{a\mu}$.

\section{$e_{a\mu}$ is proportional to $f_{a\mu}$}

Taking the condition
\begin{equation}f_{a\mu}=\rho(x) e_{a\mu}\label{eq24}\end{equation} in equation for $\omega_{\mu ab}$ (\ref{omeom})
with
\begin{align} f^{a}_{\mu}=\rho(x) e^{a}_{\mu},  && f_{a}^{\mu}=\rho(x)^{-1} e_{a}^{\mu}, &&  f^{a\mu}=\rho(x)^{-1} e^{a\mu}.
\end{align}
one obtains
\begin{align}
2\rho(x) e^{[c|}{}_{[\nu}e^{k}{}_{\sigma}\omega_{\rho]k}{}^{|d]}=2\rho(x)e^{[c}{}_{[\nu}\partial_{\rho}e^{d]}{}_{\sigma]}+e^{[c}{}_{[\nu}\partial_{\rho}\rho(x)e^{d]}{}_{\sigma]}.\label{omeomspec}
\end{align}
 To find $\omega_{\mu ab}$ we multiply (\ref{omeomspec}) with $e^{\nu}{}_{k} \eta^{di} \eta^{\rho \beta}$,  $\delta^{\beta}{}_{\delta}e^{\nu}{}_{j} e^{\rho}{}_{a} \eta^{di}$ and $e^{\beta}{}_{k} e^{\rho}{}_{a} \eta^{di}$ respectively, and solve the system of equations for $\omega_{\mu ab}$. 
\begin{equation}
\omega_{\nu ab}=\tfrac{1}{2 \rho(x)}  (e_{a\nu} e_{b}{}^{\beta} \partial_{\beta}\rho(x) -  e_{a}{}^{\beta} e_{b\nu} \partial_{\beta}\rho(x))-\tfrac{1}{2}  (e_{b}{}^{\beta} \partial_{\beta}e_{a\nu} + e_a^{\alpha}e_b^{\beta}e_{\nu}^{c}(-\partial_{\alpha}e_{c\beta}+\partial_{\beta}e_{c \alpha}) -  e_{a}{}^{\beta} \partial_{\beta}e_{b\nu} -  e_{b}{}^{\beta} \partial_{\nu}e_{a\beta} + e_{a}{}^{\beta} \partial_{\nu}e_{b\beta}).\label{eq26}
\end{equation}
This form of the $\omega_{\mu ab}$ has been expected based on the known solution from Kaku et. al \cite{Kaku:1977pa} where agreement is obtained by setting $\rho(x)$ to constant.
The condition of proportionality (\ref{eq24}) would give the action
\begin{equation}
I=\int d^4x L_s= 8\alpha \int d^4x\rho(x) (R+24 \rho(x))e \label{eq27},
\end{equation}
that is equal to Einstein gravity for $\rho(x)=1$. 
\noindent Here we used contractions \begin{align}R_{ b\mu}=R^{(0)}_{\mu\nu ab}e^{a\nu}, && R=R_{\mu a}e^{a\mu}\label{eq28}.\end{align}
Obtaining Einstein gravity from Weyl gravity has been studied from different angles \cite{Maldacena:2011mk, Metsaev:2007fq}. In \cite{Metsaev:2007fq} the relation between the Weyl and Einstein gravities have been studied via breaking conformal gauge symmetries. After imposing the relation between the gauge fields $f_{\mu\nu}$ and $e_{\mu\nu}$ which breaks the conformal gauge symmetries, the obtained Lagrangian agrees with Lagrangian in (\ref{eq27}) when $\rho(x)\rightarrow -\frac{1}{4}\rho(x_0)$, i.e. $\rho(x)$ is taken to be $-\frac{1}{4}\rho(x_0)$ constant.


\section{Discussion}

We have studied linearized equations of motion of the parity conserving action constructed from curvatures of conformal group. Since we have not imposed additional constraints by hand, the result is highly symmetric. One can notice that the symmetry which appears between the linearized fields $\chi_{\mu\nu}$ and $\theta_{\mu\nu}$ is a consequence of the symmetry which appears in the action, and speculate whether its origin reaches to the relations among the generators of special conformal transformations (SCT) and translations (T) in the conformal group.The difference between the SCTs and Ts in conformal group is due to minus sign that if absorbed in SCT generator, reemerges in changing the sign of different commutation relation. 

We have obtained the constraint equations on the fluctuations in the expansion of the gauge fields $e_{a\mu}$ and $f_{a\mu}$ around the background AdS. When the constraint equations are symmetrized one obtains two equal linearized expressions for both fields. The reason for this degeneracy beside in the conformal group, is in the perturbation around AdS space. 
For comparison, EG describes massless graviton, and CG describes one massless and one partially massless mode. Here, the perturbations are linearly dependent on each other, and system has degeneracy. In order to count precisely number of degrees of freedom one would have to perform canonical analysis of the theory. Based on current results, one may expect one massless and partially massless or massive mode.  
Inspecting the linearized equations and comparing them with the linarized equations of EG and CG, it is possible to speculate that the originating effective theory consists of the two Ricci scalars each for one metric, and additional potential. The exact form of the potential is yet to be studied.
The parameter of the theory is $\alpha$ dimensionless parameter inherited from the starting action. This is similarity of the theory with CG, but unlike in CG there are two dynamical gauge fields which is similarity with the dRGT theory.

It would be interesting to compute the observables such as one loop partition function for this theory and compare to Einstein and conformal gravity, and possibly look for generalizations to higher spins. If the generalization was to arbitrary dimensions one could consider the general d-dimensional conformal algebra and its implications which one could relate and motivate with multi-metric theories \cite{Hinterbichler:2012cn}.
One could also look into the implications of the gauge (\ref{constraint1}) and obtaining symmetric vielbeins as it was done in \cite{Deffayet:2012zc}. 

\section{Acknowledgments} 
We are grateful to Arkady Tseytlin for discussions and comments on the draft. The work was supported by the project J 4129-N27 in the framework of  Erwin-Schr\"odinger Program of the Austrian Science Fund (FWF) and by the grant ST/P000762/1 of Science and Technology Facilities Council (STFC).

\section{Appendix}

\subsection{Inverse gauge fields}

To obtain the inverse of the perturbed gauge field $f_{\mu}^a$ one starts with the  general form of the inverse gauge field $\tilde{f}^{\mu}_a$. The expansion of latter 
\begin{align}
\tilde{f}^{\mu}_b=\tilde{f}^{(0)\mu}_b+\eta\tilde{\theta}^{(1)\mu}_{b}+\eta^2\tilde{\theta}^{(2)\mu}_{b}+\eta^3\tilde{\theta}^{(3)\mu}_{b}
\end{align}
in $\mathcal{O}(0)$ order needs to satisfy $\tilde{f}^{(0)\mu}_bf_{\mu}^{(0)a}=\delta^a_b$.
Multiplication of the two expansions in the leading order gives that $\tilde{f}^{(0)\mu}_b=f^{(0)\mu}_b$. The subleading order $\mathcal{O}(1)$ gives the condition
$$ f^{(0)a}_{\alpha}\tilde{\theta}_b^{(1)\alpha}+\tilde{f}^{(0)\mu}_b\theta_{\mu}^a=0 $$ 
from which follows that $\tilde{\theta}_b^{(1)\alpha}=-f^{(0)\mu}_b\theta_{\mu}^{(1)a}f_a^{(0)\alpha}$.  The order $\mathcal{O}(2)$ leads to 
\begin{equation}
\tilde{\theta}^{(2)\alpha}_b=-f^{(0)\mu}_b\theta_{(2)\mu}^af^{(0)\alpha}_a+\theta_{\gamma}^{(1)a}f^{(0)\alpha}_af^{(0)\beta}_{b}\theta^{c}_{\beta}f^{(0)\gamma}_{c}
\end{equation}

\subsection{AdS Background}

When we consider above computation of the linear $\omega_{\mu ab}$ on the AdS background, it is most convenient to start from the equations of motion for $\omega_{\mu ab}$.
We can notice that equation (\ref{omeom}) can be written as
\begin{align}
\alpha \left( e_{c\nu}R_{\rho\sigma d}(K)+f_{c\nu}R_{\rho\sigma d}(P) \right)\epsilon^{\mu\nu\rho\sigma}\epsilon^{abcd}=0 \label{kpom}
\end{align}
for 
\begin{align}
R_{\mu\nu a}(P)&=-(\partial_{\mu}e_{a\nu}-\omega_{\mu a}^be_{b\nu})+(\partial_{\nu}e_{a\mu}-\omega_{\nu a}^b e_{b\mu})
 \label{curvs2} \\
R_{\mu\nu a}(K)&=-(\partial_{\mu}f_{a\nu}-\omega_{\mu a}^bf_{b\nu})+(\partial_{\nu}f_{a\mu}-\omega_{\nu a}^bf_{b\mu})
\label{curvs3}.
\end{align}
In the leading order (\ref{kpom}) reads 
\begin{align}
\alpha \left( v^{}_{c\nu}R^{(0)}_{\rho\sigma d}(K)+f^{(0)}_{c\nu}R^{(0)}_{\rho\sigma d}(P) \right)\epsilon^{\mu\nu\rho\sigma}\epsilon^{abcd}=0
\end{align}
where we have used index  $(0)$ in $R^{(0)}_{\mu\nu a}$ to accent the order of perturbation. 
Since we use $f^{(0)}_{c\nu}=v^{}_{c\nu}$ the equation reduces to
\begin{align}
2\alpha  v^{}_{c\nu}R^{(0)}_{\rho\sigma d}(P) \epsilon^{\mu\nu\rho\sigma}\epsilon^{abcd}=0
\end{align}
where we can recognize the appearance of the no torsion condition, which corresponds to the requirement that the covariant derivative of the AdS vielbein vanishes. 
That means in the subleading order 
\begin{align}
\alpha \left[ v^{}_{c\nu}(R^{(1)}_{\rho\sigma d}(K)+R^{(1)}_{\rho\sigma d}(P)) +\chi^{}_{c\nu}R^{(0)}_{\rho\sigma d}(P)+\theta^{}_{c\nu}R^{(0)}_{\rho\sigma d}(P)   \right]\epsilon^{\mu\nu\rho\sigma}\epsilon^{abcd}=0
\end{align}
the second and the third term may be taken to zero due to no torsion condition so one obtains 
\begin{align}
\alpha  v^{}_{c\nu}(R^{(1)}_{\rho\sigma d}(K)+R^{(1)}_{\rho\sigma d}(P))  \epsilon^{\mu\nu\rho\sigma}\epsilon^{abcd}=0\label{eq100}
\end{align}
for
\begin{align}
R^{(1)}_{\rho\sigma d}(P)=-(\partial_{\mu}\chi^{}_{a\nu}-\omega_{\mu }^{(0)b}{}_a\chi^{}_{b\nu}-\omega_{\mu }^{(1)b}{}_{a}v^{}_{b\nu})+\partial_{\nu}\chi^{}_{a\mu}-\omega_{\nu }^{(0)b}{}_a\chi^{}_{b\mu}-\omega_{\nu }^{(1)b}{}_{a}v^{}_{b\mu},\label{eq101}
\end{align}
and $R^{(1)}_{\rho\sigma d}(K)$ gives the same expression with $\theta^{}_{a\mu}$ on the place of $\chi^{}_{a\mu}$ in (\ref{eq101}).

Analogously to the procedure for equation (\ref{omeom}) we can dualize (\ref{eq100}) to obtain the equation for the $\omega^{(1)}_{\mu ab}$ 
\begin{align}
 v^{}_{c[\nu}(R_{\rho\sigma] d}^{(1)}(K)+R_{\rho\sigma] d}^{(1)}(P)- v^{}_{d[\nu}(R_{\rho\sigma] c}^{(1)}(K)+R_{\rho\sigma] c}^{(1)}(P)) =0. \label{eq102}
\end{align}
To solve the equation (\ref{eq102}) for the $\omega^{(1)}_{\mu ab}$ we 
obtain three tensorial equation whose manipulation leads to the expression for $\omega^{(1)}_{\mu ab}$. The simplification that can be taken for AdS background, is that AdS background is Weyl flat and one can define 
\begin{align}
v^{}_{a\mu}=\rho(x)\delta_{a\mu}.\label{adscon}
\end{align}
Here $\rho(x)$ denotes function of the coordinates on the manifold. 
 The multiplication for obtaining the tensorial equations is therefore also done by using (\ref{adscon}). To express $\omega^{(1)}_{\mu ab}$ we use Mathematica package xAct \cite{xact} and classify the terms as
\begin{enumerate}
\item terms $\omega^{(1)}_{\mu ab}(\omega, \chi^{},\theta^{})$ with $\omega_{\mu ab}$, $\chi_{a\mu}^{(1)}$  and $\theta_{a\mu}^{(1)}$
\item terms $\omega^{(1)}_{\mu ab}(\partial \chi^{})$ with $\partial_{\mu}\chi^{}_{a\nu}$
\item terms $\omega^{(1)}_{\mu ab}(\partial \theta^{})$ with $\partial_{\mu}\theta^{}_{a\nu}$.
\end{enumerate}
There is no terms that involve the partial derivative acting on the background vielbein. The reason for this becomes clear from equation (\ref{eq101}). In the linear order we can have the partial derivative of background vielbein only from $\omega^{(0)}_{\mu ab}$, while the remaining terms vanished due to no torsion condition. (Below we omit writing $(0)$ in $\omega^{(0)}$ for simplicity.)

For the terms 1. we obtain 
\begin{align}
\tilde{\omega}^{(1)}_{\kappa ak}(\omega, \chi^{},\theta^{})=-\frac{1}{4\rho}\left[(\omega_{k}{}^{b}{}_{\kappa} + \omega_{\kappa}{}^{b}{}_{k}) (\theta^{}_{ba}+\chi^{}_{ba}) -  \omega_{a}{}^{b}{}_{k} (\theta^{}_{b\kappa}+\chi^{}_{ba})\right]\label{eq104}
\end{align}
here 
\begin{align} \omega^{(1)}_{\kappa ak}(\omega, \chi^{},\theta^{})=\tilde{\omega}^{(1)}_{\kappa ak}(\omega, \chi^{},\theta^{})-\tilde{\omega}^{(1)}_{\kappa ka}(\omega, \chi^{},\theta^{}).\end{align}
Terms 2. are $\omega^{(1)}_{\mu ab}(\partial \chi^{})=\tilde{\omega}^{(1)}_{\mu ab}(\partial \chi^{})-\tilde{\omega}^{(1)}_{\mu ba}(\partial \chi^{})$
\begin{align}
\tilde{\omega}^{(1)}_{\mu ab}(\partial \chi^{})=- \tfrac{1}{4\rho} \delta_{k}{}^{\alpha} \delta_{\kappa}{}^{b} \partial_{a}\chi^{}_{b\alpha} -  \tfrac{1}{4\rho} \partial_{a}\chi^{}_{k\kappa} -  \tfrac{1}{4\rho} \delta_{k}{}^{\alpha} \partial_{\kappa}\chi^{}_{a\alpha},
\end{align}
and terms 3. are equal to terms 2. with $\theta^{}_{a\mu}$ on the place of $\chi^{}_{a\mu}$: $\omega^{(1)}_{\mu ab}(\partial \theta^{})=\tilde{\omega}^{(1)}_{\mu ab}(\partial \theta^{})-\tilde{\omega}^{(1)}_{\mu ba}(\partial \theta^{})$
\begin{align}
\tilde{\omega}^{(1)}_{\mu ab}(\partial \theta^{})=- \tfrac{1}{4\rho} \delta_{k}{}^{\alpha} \delta_{\kappa}{}^{b} \partial_{a}\theta^{}_{b\alpha} -  \tfrac{1}{4\rho} \partial_{a}\theta^{}_{k\kappa} -  \tfrac{1}{4\rho} \delta_{k}{}^{\alpha} \partial_{\kappa}\theta^{}_{a\alpha}.\label{eq107}
\end{align}
To identify the covariant derivatives let us rewrite the $\theta^{}_{a\mu}$ part of the equation (\ref{eq104}) with indices on $\omega_{\mu ab}$ not contracted, equation (\ref{eq104}) 
\begin{align}
-\frac{1}{4\rho}\left[(\delta_k^{\beta}\delta_a^{\alpha}\delta_{\kappa}^{c}\omega_{\beta}{}^{b}{}_{c} + \delta^{\alpha}_a \omega_{\kappa}{}^{b}{}_{k}) \theta^{}_{b\alpha} - \delta_a^{\alpha} \omega_{\alpha}{}^{b}{}_{k} \theta^{}_{b\kappa}\right]\label{eq108}
\end{align}
Combining the third term from the (\ref{eq108}) and the third term from (\ref{eq104}) we have 
\begin{align}
\delta^{\alpha}_a(\partial_{\alpha}\theta_{k\kappa}-\omega_{\alpha}{}^c{}_{k}\theta_{c\kappa})=\delta^{\alpha}_a\nabla_{\alpha}\theta_{k\kappa}.
\end{align}
The remaining terms from (\ref{eq108}) analogously combine with the antisymmetric pairs of the terms in (\ref{eq104}) into covariant derivatives. Taking into account $\chi^{}_{a\mu}$, $\theta^{}_{a\mu}$ and equations (\ref{eq104}) to (\ref{eq107}) we obtain
\begin{align}
\omega^{(1)}_{\kappa ak}=-\frac{1}{4\rho}\left( \delta^{\alpha}_a\nabla_{\alpha}(\theta_{k\kappa}+\chi_{k\kappa})+\delta^{\alpha}_k\nabla_{\kappa}(\theta^{}_{a\alpha}+\chi^{}_{a\alpha})+\delta^{\alpha}_k\delta^b_{\kappa}\delta^{\beta}_a\nabla_{\beta}(\theta^{}_{b\alpha}+\chi^{}_{b\alpha}) \right)-a \leftrightarrow k\label{omegap}
\end{align}

For the Einstein gravity (EG) spin connection holds
\begin{align}
\omega^{EG}{}_{\mu}{}^a{}_b=-e_{b}{}^{\nu}\mathcal{D}_{\mu}e_{a}{}^{\mu} \label{egrel}
\end{align}
which is equal to the (\ref{specom}) in the leading order. Where we denote covariant derivative with $\mathcal{D}$. 
In the linearized order this is
\begin{align}
\omega^{EG(1)}{}_{\mu}{}^a{}_{b}(\chi)=-\tilde{\chi}_{b}{}^{\nu}\mathcal{D}_{\mu}v_{a}{}^{\nu} v_{b}{}^{\nu}\mathcal{D}^{(1)}_{\mu}v_{a}{}^{\nu} -v_{b}{}^{\nu}\mathcal{D}_{\mu}\tilde{\chi}_{a}{}^{\nu}.\label{eq57vn}
\end{align}
We can write (\ref{omegap1}) as 
\begin{align}
\omega^{(1)}_{\mu ab}(\chi+\theta)=\omega^{EG(1)}_{\mu ab}(\chi)+\omega^{EG(1)}_{\mu ab}(\theta) \label{sumeg},
\end{align}
linearizing (\ref{sumeg}) around AdS we can write the terms  
\begin{align}
\omega^{(1)EG}_{\mu ab (AdS)}(\chi)&=\frac{1}{2 \rho^2}(\bigl(\chi^{}_{b\mu} -  \chi^{}_{\mu b}\bigr) \partial_{a}\rho + \bigl(- \chi^{}_{a\mu} + \chi^{}_{\mu a}\bigr) \partial_{b}\rho + \bigl(- \chi^{}_{ab} + \chi^{}_{ba}\bigr) \partial_{\mu}\rho \\&+ \bigl((-\eta_{b\mu} -  \eta_{\mu b}) \chi^{\nu}{}_{a} + (\eta_{a\mu} + \eta_{\mu a}) \chi^{\nu}{}_{b}\bigr) \partial_{\nu}\rho)\\
&+\frac{1}{2\rho}(\partial_a\chi^{}{}_{b\mu}-\partial_{b}\chi^{}{}_{a\mu}+\delta_b{}^{\nu}\partial_{\mu}\chi^{}{}_{a\nu}-\delta_a{}^{\nu}\partial_{\mu}\chi^{}{}_{b\nu}+\delta_{\mu}{}^c(\delta_b{}^{\lambda}\partial_{a}\chi^{}{}_{c\lambda}-\delta_a{}^{\lambda}\partial_b\chi^{}{}_{c\lambda}))
\label{wegads}
\end{align}
and $\omega^{(1)EG}_{\mu ab (AdS)}(\theta)$ analogously.
We can notice that choice of symmetric perturbation $\chi_{\mu b}=\chi_{b\mu}$, $\chi_{ab}=\chi_{ba}$ reduces  (\ref{wegads}) to 
\begin{align}
\omega^{(1)EG}_{\mu ab (AdS)}{}_{\text{symmetric}}&=\frac{(- \eta_{\mu b} \chi_{a}{}^{\nu} +  \eta_{\mu a} \chi_{b}{}^{\nu}) \partial_{\nu}\rho}{ \rho^2}
\\& +\frac{1}{2\rho}(\partial_a\chi^{}{}_{b\mu}-\partial_{b}\chi^{}{}_{a\mu}+\delta_{\mu}{}^c(\delta_b{}^{\lambda}\partial_{a}\chi^{}{}_{c\lambda}-\delta_a{}^{\lambda}\partial_b\chi^{}{}_{c\lambda})).
\end{align}
(\ref{sumeg}) also requires that
\begin{equation}\scalebox{0.5}
-\frac{1}{4\rho}\left( v^{\alpha}_a\nabla_{\alpha}(\chi^{}_{k\kappa})+v^{\alpha}_k\nabla_{\kappa}(\chi^{}_{a\alpha})+v^{\alpha}_kv^b_{\kappa}v^{\beta}_a\nabla_{\beta}(\chi^{}_{b\alpha}) \right)-a \leftrightarrow k=(-\tilde{\chi}_{k}{}^{\nu}\mathcal{D}_{\kappa}v_{\nu}{}^d+v^{\nu}{}_k\Gamma^{(1)\alpha}{}_{\kappa\nu}v_{\alpha}{}^d-v^{\nu}{}_k\mathcal{D}_{\kappa}\chi_{d}{}^{\nu})\eta_{dc}\label{vielmet}
\end{equation}
where $\Gamma^{(1)\alpha}_{\kappa\nu}$ is Christoffel $\Gamma^{\alpha}_{\kappa\nu}=\frac{1}{2}e^{\alpha\beta}(\partial_{\kappa}e_{\beta\nu}+\partial_{\nu}e_{\beta\kappa}-\partial_{\beta}e_{\kappa\nu})$ expanded
 for $e_{\mu\nu}=e_{a\mu}e^{a}{}_{\nu}$, its expansion $e_{\mu\nu}=v^{}_{\mu\nu}+h_{\mu\nu}$, and \begin{equation}h_{\mu\nu}=v^{}_{a\mu}\chi^{a}{}_{\nu}+\chi^{}_{a\mu}v^{a}{}_{\nu}.\end{equation} We have defined $h_{\mu\nu}$ as symmetric term in perturbation of the $e_{\mu\nu}$. Expansion is analogous for $\theta_{a\mu}$, 
\begin{align}
q_{\mu\nu}=v_{a\mu}\theta^a{}_{\nu}+\theta_{a\mu}v^a{}_{\nu}.\label{qdef}
\end{align}

Proving that (\ref{sumeg}) holds 
makes possible writing the perturbation as a sum of perturbations in Einstein gravity.
We can consider the linearized projection of the Riemann tensor from the vielbein to metric formulation. 
For the projection of the Riemann tensor we know $R^{\lambda}{}_{\sigma\mu\nu}=e_a{}^{\lambda}e^b{}_{\sigma}R_{\mu\nu}{}^a{}_b$. When we rewrite definition of $R^{a}{}_{b\mu\nu}$ (12) in terms of the (\ref{egrel}) $\omega_{\mu}{}^a{}_b=e^a{}_{\alpha}e_b{}^{\beta}\Gamma^{\alpha}_{\mu\beta}-e_b{}^{\alpha}\partial_{\mu}e^a{}_{\alpha}$ the projection gives us $R^{\lambda}_{\sigma\mu\nu}$. The terms in the computation that contain one partial derivation $\partial_{\mu}$, $\partial_{\nu}$ and their combination $f(\partial_{\mu},\partial_{\nu})$ (for $f$ function in $\partial_{\mu}$ and $\partial_{\nu}$) in the leading order separately cancel. Analogously, we consider them in linearized order. 

We write the projection 
\begin{align}
R^{\lambda}{}_{\sigma\mu\nu}&=e^{\lambda}_ae_{\sigma}^b(-\partial_{\mu}(e_{\rho}^ae^{\tau}_b\Gamma^{\rho}_{\nu\tau})+\partial_{\nu}(e_{\rho}^ae^{\tau}_b\Gamma^{\rho}_{\mu\tau})+\partial_{\mu}e^{\tau}_b\partial_{\nu}e_{\tau}^a-\partial_{\nu}e^{\tau}_b\partial_{\mu}e_{\tau}^a\\&-(e_{\rho}^ae^{\tau}_c\Gamma_{\mu\tau}^{\rho}-e^{\tau}_c\partial_{\mu}e_{\tau}^a)(e_{\rho'}^ce^{\tau'}_b\Gamma_{\nu\tau'}^{\rho'}-e^{\tau'}_b\partial_{\nu}e_{\tau'}^c)\\&+(e_{\rho}^ae^{\tau}_c\Gamma_{\nu\tau}^{\rho}-e^{\tau}_c\partial_{\nu}e_{\tau}^a)(e_{\rho'}^ce^{\tau'}_b\Gamma_{\mu\tau'}^{\rho'}-e^{\tau'}_b\partial_{\mu}e_{\tau'}^c))
\end{align} 
and linearize it. 
In the linearized order projection is
\begin{align}
R^{(1)\lambda}{}_{\sigma\mu\nu}(\chi)=v^{a\lambda }v^{b}_{\sigma} R^{(1)}{}_{\mu\nu ab}(\chi)+R^{(0)}{}_{\mu\nu ab}v^{a\lambda }\chi_{\sigma}^b+R^{(0)}{}_{\mu\nu ab}\tilde{\chi}^{a\lambda }v^{b}_{\sigma}\label{eq89}
\end{align}
the subleading order of
$
R^{\lambda}_{\sigma\mu\nu}=-\partial_{\mu}\Gamma^{\lambda}_{\nu\sigma}+\partial_{\nu}\Gamma^{\lambda}_{\mu\sigma}-\Gamma^{\lambda}_{\mu\alpha}\Gamma^{\alpha}_{\nu\sigma}+\Gamma^{\lambda}_{\nu\alpha}\Gamma^{\alpha}_{\mu\sigma}
$
\subsection{Comparison with Einstein gravity}

Analogous consideration of Einstein gravity would lead to equations of motion in the subleading order 
\begin{align}
G_{a\mu}^{(1)}=R_{a\mu}^{(1)}-\frac{1}{2}R^{(1)}e_{a\mu}-\frac{1}{2}R\chi_{a\mu}=0.
\end{align}
Using the above method and De Donder gauge leads to the constraint on $\chi_{a\mu}$
\begin{align}
-\tilde{\lambda}\chi_{\mu\nu}-\mathcal{D}^2\chi_{\mu\nu}+\frac{1}{2}(2\tilde{\lambda}\chi+\mathcal{D}^2\chi)v_{\mu\nu}=0.\label{einst}
\end{align}
To compare this with the familiar result for the linearized Einstein operator we have to consider $h_{\mu\nu}=2\chi_{\mu\nu}$ which is symmetric and
\begin{align}G^{(1)}_{\mu\nu}=G_{a\mu}^{(1)}v^{a}_{\nu}+G^{(0)}_{a\mu}\chi^{a}_{\nu},\end{align} where $G^{(0)}_{a\mu}\chi^{a}_{\nu}=-3\tilde{\lambda}\chi_{\mu\nu}$. We also need to take into account cosmological constant which is $6\tilde{\lambda}\chi_{\mu\nu}$ for four dimensions. Adding this to (\ref{einst})
we obtain familiar result
\begin{align}
2\tilde{\lambda}\chi_{\mu\nu}-\mathcal{D}^2\chi_{\mu\nu}+\frac{1}{2}(2\tilde{\lambda}\chi+\mathcal{D}^2\chi)v_{\mu\nu}=0.
\end{align}

\subsection{Relations used in text}
\begin{align}
\frac{1}{4}\epsilon_{abcd}\epsilon^{\mu\nu\rho\sigma}e^a{}_{\mu}e^b{}_{\nu}e^c{}_{\rho}e^d{}_{\sigma}&=e \\
\frac{1}{2!}\epsilon_{abcd}\epsilon^{\mu\nu\rho\sigma}e^c_{\rho}e^d_{\sigma}&=e(e_{a}^{\mu}e_b^{\nu}-e_a^{\nu}e_b^{\mu})\\
\delta e&=e e_a^{\mu}\delta e^a_{\mu} 
\end{align}

\bibliographystyle{apsrev}
\bibliography{bibliography}

\end{document}